\documentclass[aps,showpacs,amssymb,dvipsnames,sort&compress,numbers,sort&compress,nofootinbib,superscriptaddress,twocolumn,longbibliography,floatfix]{revtex4-1}
\pdfoutput=1
\usepackage[utf8]{inputenc}
\usepackage[english]{babel}
\usepackage[T1]{fontenc}
\usepackage{amsmath}
\usepackage{amssymb}
\usepackage{physics}
\usepackage{tikz}
\usepackage{lipsum}

\begin{document}

\title{Genuine multipartite correlations in Dicke Superradiance}
\date{\today}
\author{Susane Calegari}
\email{calegari.su@gmail.com}
\affiliation{Departamento de Física, Universidade Federal de Santa Catarina, CEP 88040-900, Florianóplis, SC, Brazil}
\author{Antônio C. Lourenço}
\affiliation{Departamento de Física, Universidade Federal de Santa Catarina, CEP 88040-900, Florianóplis, SC, Brazil}
\author{Gabriel T. Landi}
\affiliation{Instituto de Física, Universidade de São Paulo, CEP 05314-970, São Paulo, São Paulo, Brazil}
\author{Eduardo I. Duzzioni}
\affiliation{Departamento de Física, Universidade Federal de Santa Catarina, CEP 88040-900, Florianóplis, SC, Brazil}

\begin{abstract}
A thorough understanding of the structure of correlations in multipartite systems is essential for the success of most quantum information processing applications. 
The problem, however,  quickly becomes non-trivial as the size of the multipartition increases.
With this motivation in mind, in this paper, we put forth a detailed study of genuine multipartite correlations (GMCs) in the Dicke model of superradiance. We compute all genuine $k$-partite correlations for Dicke states with arbitrary excitations and use these results to characterize the evolution of multipartite correlations during the superradiant dynamics. Non-trivial effects in the way correlations in Dicke states are distributed between the multiple parts are found, showing strong finite-size effects. We also employ the concept of weaving to classify how multipartite correlations scale with the number of particles in the system.
\end{abstract}

\maketitle

\section{Introduction}
The ability to coherently manipulate quantum correlations in the laboratory is one of the main drives behind emerging quantum technologies. 
As soon as one moves beyond the bipartite paradigm, however, the complexity of classifying multipartite correlations increases rapidly. For instance, even three qubits can already be entangled in inequivalent ways \cite{PhysRevA.62.062314,coffman2000distributed}.
Increasing the number of qubits further makes the situation impractical to characterize.

Among the several questions one has to address in the multipartite scenario, one of particular importance is how to define and characterize \emph{genuine} multipartite correlations (GMCs). Given a $N$-partite quantum state, the GMCs of order $k\le N$ represent the amount of information that cannot be obtained from any cluster of size smaller than $k$. For instance, the quantum correlations in a GHZ state \cite{greenberger1990bell} of $N$ qubits is genuinely $N$-partite, whereas the $N/2$ product of Bell pairs will only have bipartite correlations. In recent years there have been substantial efforts to characterize the GMCs of a variety of quantum states and scenarios \cite{toth2005detecting,walczak2010information,grudka2008note,giorgi2013genuine,zhang2012two,bai2013exploring,grimsmo2012dynamics,li2012detecting,giorgi2011genuine,giorgi2013erratum,maziero2012genuine,aolita2012fully,novo2013genuine,mendoncca2015heuristic,girolami2017quantifying}.

In particular, Girolami~\emph{et.~al.} \cite{girolami2017quantifying}  recently introduced a measure for (quantum plus classical) GMCs of order $k\le N$ based on the general framework of distance-based information-theoretic quantifiers (c.f. Ref.~\cite{modi2010unified}).
Their measure satisfies the postulates put forth in~Ref.~\cite{bennett2011postulates}, which are expected to hold for any valid measure of genuine multipartite correlations. It also satisfies the criteria of monotonicity under local operations expected for systems invariant under subsystems permutations. Moreover, the measure is based on the quantum relative entropy and thus has the advantage of being computationally more feasible than other distance measures.

Calculations of GMCs of order $k$ usually require accessing all possible partitions of a state, which quickly becomes prohibitive as the number of components increases. For this reason, studies on multipartite correlations have been restricted mostly to few-body systems or highly symmetric states, such as GHZ-like  \cite{girolami2017quantifying,susa2018weaving} or Dicke states \cite{novo2013genuine,i2016characterizing}.
The calculations, in this case, simplify dramatically due to permutation invariance. 

Dicke states offer a nice example of how multipartite correlations may affect relevant physical processes in controlled quantum systems. 
Superradiance is a coherent radiative phenomenon resulting from atomic cooperativeness, where the atomic ensemble spontaneously emits radiation in a shorter amount of time \cite{dicke1954coherence,gross1982superradiance}.
Dicke states are known to be highly entangled \cite{Bergmann_2013,moreno2018all}. 
During the superradiant emission, however, all Dicke states are incoherently mixed, so that the entanglement between any two partitions is null \cite{wolfe2014certifying,yu2016separability,tura2018separability} and the remaining correlations are either classical or discord-like \cite{santos2016}.
Beyond that, to the best of our knowledge, practically nothing else is known about genuine correlations patterns in Dicke superradiance. Actually, for higher-dimensional systems, the problem of identifying these correlations is an NP-hard problem about which very little is known.

In this work, we address how genuine $k$-partite correlations behave during the superradiant emission. 
For this, we employ the framework developed in Ref. \cite{girolami2017quantifying} to study GMCs in pure Dicke states with an arbitrary number of excitations and then use these results to study the evolution of GMCs during the superradiant emission dynamics.
In addition to the set of all $k$-partite correlations, we also use the concept of weaving \cite{susa2018weaving} to quantify how the correlations scale with different system sizes. 

This paper is divided as follows. The GMCs formalism of Ref.~\cite{girolami2017quantifying} is reviewed in Sec.~\ref{sec:gmc} and then applied to the set of Dicke states in Sec.~\ref{sec:dicke}. 
In Sec.~\ref{sec:dynamics} we move on to describe the time evolution of GMCs during the superradiant dynamics. 
Conclusions are summarized in Sec.~\ref{sec:conc}. 
Finally, in Appendix~\ref{ap: Dicke} we provide expressions for the partial trace of a general Dicke state, as well as an incoherent mixture of Dicke states and its von Neumann entropy.

\section{Quantifying genuine multipartite correlations}\label{sec:gmc}
In this section, we briefly review the measure of GMCs and weaving introduced in Ref.~\cite{girolami2017quantifying}.
The measure can be defined for any distance quantifier, but it is simplified significantly if one uses the quantum relative entropy, as will be done here.
We consider a finite dimensional $N$-partite quantum system described by a certain density matrix $\rho_N$ and let $P_k$ denote the set of all marginalizations of $\rho_N$ having clusters of at most size  $k$; viz., $P_k = \{\bigotimes_{i=1}^m \rho_{k_i}, \;\sum_{i=1}^m k_i = N, \;k = \max k_i \} $. 
The total amount  GMCs which have order higher than $k$ is then defined as the smallest distance between $\rho_N$ and $P_k$:
\begin{equation}
    S^{k \to N}(\rho_N) := \underset{\sigma_N\in P_k}{\text{min}} S(\rho_N||\sigma_N),
    \label{eq:Shigherkgenuine}
\end{equation}
where $S(\rho||\sigma) = \tr(\rho \ln \rho - \rho \ln \sigma)$ is the quantum relative entropy.
Particularly important, are the total correlations, which is the distance between $\rho_N$ and the maximally marginalized state \cite{modi2010unified}:
\begin{equation}
T(\rho_N):=S^{1\to N}(\rho_N).
\end{equation}
This quantity measures the total amount of correlations present in the global state, which is lost if one only has access to the reduced states of each part. 

From $S^{k\to N}$ one may then define the genuine $k$-partite correlations as those which are present in a $k$-partition but absent in a $(k-1)$-partition, i.e.,
\begin{equation}
    S^k(\rho_N):= S^{k-1\to N}(\rho_N)-S^{k\to N}(\rho_N).
    \label{eq:kgenuine}
\end{equation}
This quantity measures the amount of correlations which are genuinely of order $k$.

In this paper we shall restrict to permutationally invariant subsystems. In this case, the $k$-partite state in  $P_k$ which minimizes $S(\rho_N||\sigma_N)$ is
\begin{equation}
    \sigma_N =\left( \bigotimes_{i=1}^{\lfloor N/k\rfloor}\rho_k\right)\otimes\rho_{N\text{mod}k},
\end{equation}
where $\rho_k = \Tr_{N-k}\rho_N$ is the reduced matrix of a cluster of $k$ parts and $\lfloor x \rfloor$ is the floor function\cite{modi2010unified,szalay2015}.
Writing $S(\rho_N || \bigotimes_{i=1}^m \rho_{k_i}) = \sum_{i=1}^m S(\rho_{k_i})- S(\rho_N)$, the GMCs of order higher than $k$ simplify to 
\begin{align}\label{eq:Skhigher_symm}
    S^{k\to N}(\rho_N)=&\left\lfloor \frac{N}{k} \right\rfloor S(\rho_k)-S(\rho_N)\\
    &+(1-\delta_{N \text{mod} k,0})S(\rho_{N\text{mod} k}),\nonumber
\end{align}
which is much more tractable from a computational point of view.

To rank GMCs through a single index, the authors in Ref.~\cite{girolami2017quantifying} also introduced the concept of weaving, as the weighted sum of genuine multipartite correlations,
\begin{align}
    W_S(\rho_N)&=\sum_{k=2}^N\omega_k S^k(\rho_N)
    = \sum_{k=1}^{N-1} \Omega_k S^{k\to N}(\rho_N),
    \label{eq:weaving}
\end{align}
where $\omega_k = \sum_{i=1}^{k-1}\Omega_i$ and $\omega_k \in\mathbb{R}^+$. The choice of the weights determines the meaning of the weaving measure. For example, the total correlations can be measured with $\omega_k = 1$, $\forall k$, and the genuine $l$-partite correlations with $\omega_k = \delta_{k l}$, $\forall k$. To quantify how the GMCs scale with the system size $N$ one may use $\omega_k=k-1$ or $\Omega_k = 1$, $\forall k$.

\section{\label{sec:dicke}GMCs in Dicke states}

The Dicke states $\ket{N,n_e}$, representing $n_e$ out of $N$ qubits in the excited state,  are defined as 
\begin{equation}
    \ket{N,n_e}=\frac{1}{\sqrt{\binom{N}{n_e}}} \sum_i \mathcal{P}_i \left(\ket{1}^{\otimes n_e} \ket{0}^{\otimes(N-n_e)}\right),
    \label{eq:DickeState2}
\end{equation}
where $\binom{N}{n_e}$ is the binomial coefficient and the sum is over all possible permutations that lead to different states ${\mathcal{P}_i}$.
The $N+1$ Dicke states span only the symmetric subspace of the full Hilbert space of $N$ qubits. 
In general, they are highly entangled \cite{Bergmann_2013,moreno2018all}, except for $|N,0\rangle$ and $|N,N\rangle$. 

\begin{figure}[h!]
    \centering
    \includegraphics[width=0.5\textwidth]{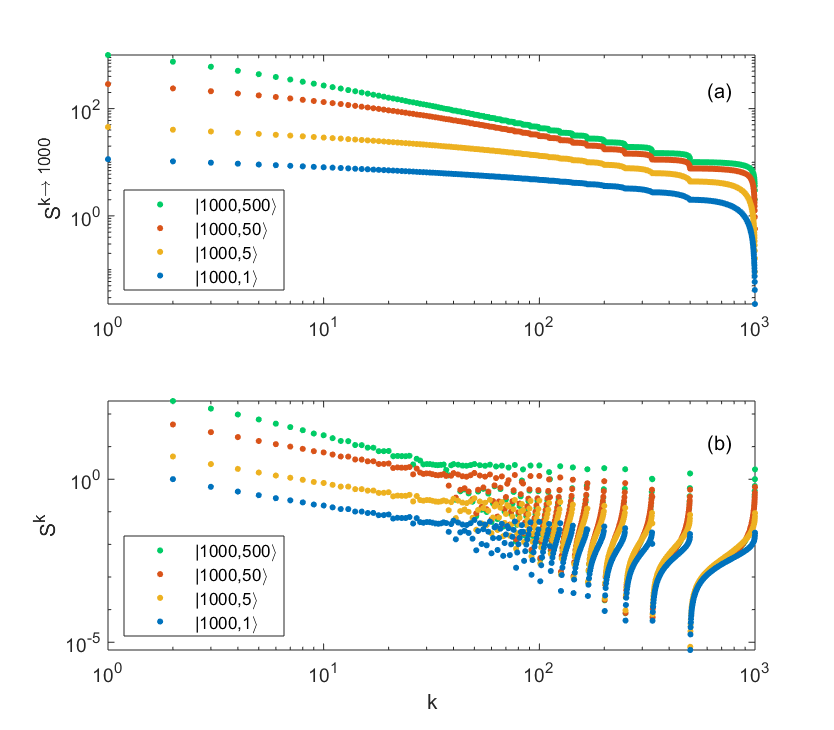}
    \caption{GMCs for Dicke states with 1, 5, 50, and 500 excitations. The GMCs (a) of order higher than $k$, $S^{k \to 1000}$, and (b) genuine $k$-partite correlations, $S^k$, increase as the number of excitations in Dicke states increases for every $k$-partitions until the maximum value which occurs for $n_e=N/2$.}
    \label{fig:dicke}
\end{figure}

For the Dicke states~(\ref{eq:DickeState2}) it is possible to find closed-form expressions for $S^{k\to N}$  (\ref{eq:Skhigher_symm}) and $S^k$ (\ref{eq:kgenuine}).
The calculations are given in appendix  \ref{ap: Dicke} and the results are plotted in Fig. (\ref{fig:dicke}). We present both  $S^{k\to N}$ and $S^k$~vs.~$k$ for $N=1000$ particles and $n_e=1,5,50$, and $500$ excitations. As the states $\ket{N,n_e}$ and $\ket{N,N-n_e}$ have the same amount of correlations, it suffices to consider $n_e \leq N/2$. 

\begin{figure}[htbp]
    \centering
    \includegraphics[width=0.5\textwidth]{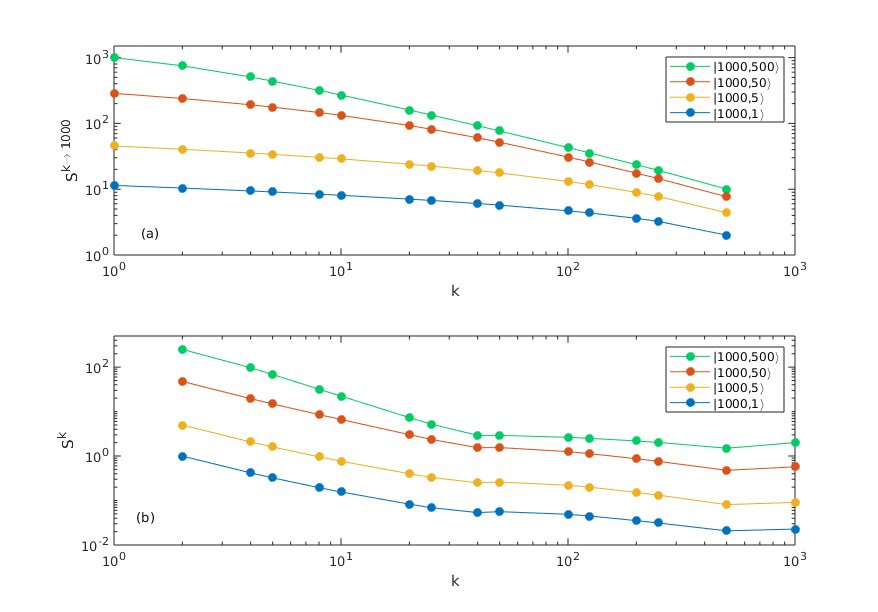}
    \caption{GMCs for Dicke states with 1, 5, 50, and 500 excitations for $k$ satisfying $\mod(N,k)=0$. The GMCs (a) of order higher than $k$, $S^{k \to 1000}$, and (b) genuine $k$-partite correlations, $S^k$, increase with  the number of excitations for every $k$-partition until the maximum value which occurs for $n_e=N/2$.}
    \label{fig:dist_modzero}
\end{figure}

We see in Fig. (\ref{fig:dicke}) that both GMCs increase with the number of excitations $n_e$, with $\ket{N,N/2}$ having the largest correlations overall. The quantity $S^{k\to N}$ in Fig. (\ref{fig:dicke}a) is seen to be monotonically decreasing with $k$, as it should be, since it represents the total distance between the $N$-body state and a $k$-partition. For small values of $k$, its decrease is smooth. 
But as $k$ becomes comparable with $N$, abrupt jumps are observed. These sudden changes in the value of the GMCs can be due to the floor function $\lfloor k/N \rfloor$ for $k$ in the same order of magnitude of $N$. Focusing on the partitions on with $\mod(N,k)=0$, we observe a smooth decrease of $S^{k\to N}$, see Fig. (\ref{fig:dist_modzero}a).
On the other hand, $S^{k}$ [Eq.~(\ref{eq:kgenuine})] measures the genuine $k$-partite correlations and therefore does not have to be monotonic. 
As seen in Fig. (\ref{fig:dicke}b), the largest GMCs occur for $k=2$. For small $k$, $S^k$ decreases smoothly, whereas for $k \gtrsim 50$,  finite size effects cause $S^k$ to fluctuate considerably. Fig. (\ref{fig:dist_modzero}b) shows that even considering only partitions in which $\mod(N,k)=0$, we continue to see non-monotonic behavior in $S^k$, e.g. $S^{40}>S^{50}$, which makes us believe that this is a physical behavior.

\begin{figure}[htbp]
    \centering
    \includegraphics[width=0.5\textwidth]{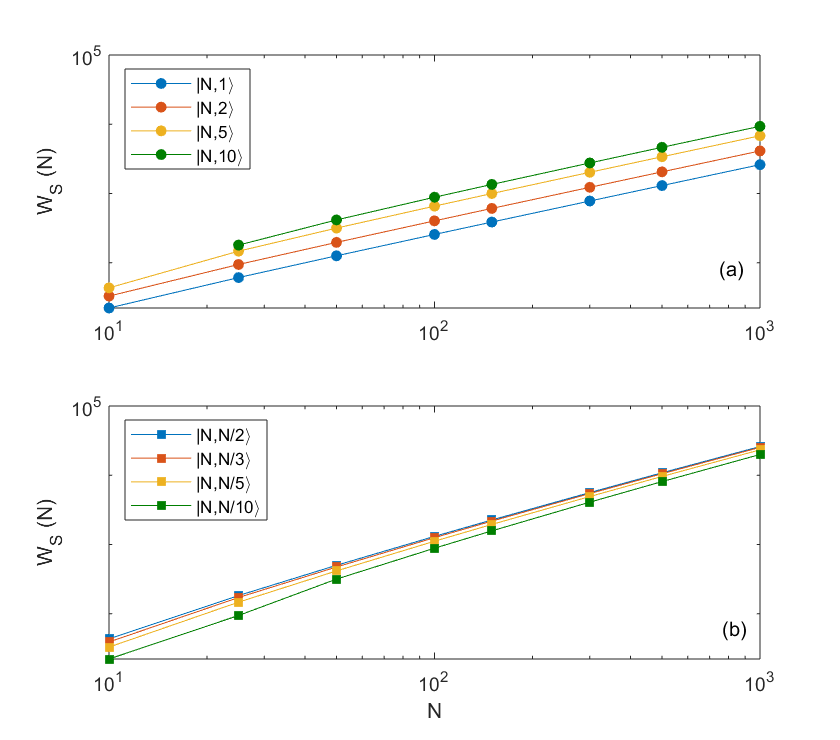}
    \caption{The scalability of the correlations measured by weaving for two classes of Dicke states: in one class (a) the number of excitations (1, 2, 5, and 10) in the state is independent of $N$, while in the other one (b) the number of excitations is a fraction of $N$ ($N/10$, $N/5$, $N/3$, and $N/2$).
    }
    \label{fig:dicke_weaving2}
\end{figure}
The weaving~(\ref{eq:weaving}) (with $\omega_k = k-1$) is shown in Fig.~(\ref{fig:dicke_weaving2}) as a function of $N$. 
We consider two scenarios. In the first one, we analyze $W(N)$ for a fixed number of excitations,  ($n_e=1$, 2, 5, and 10). 
In the other one, we analyze $W(N)$ when $n_e$ scales with $N$ as  $n_e=N/10$, $N/5$, $N/3$, and $N/2$. In both cases, the weaving behaves almost linear in the Log-Log scale, but in Fig.~(\ref{fig:dicke_weaving2}b) the weaving increases faster than in Fig.~(\ref{fig:dicke_weaving2}a). This result is in agreement with the fact that the genuine correlations in Dicke states increase with the number of excitations $n_e$, but limited to the maximum value of correlations determined by the state with $n_e=N/2$, as explained above. 

\section{\label{sec:dynamics}Dicke Superradiance}

\subsection{Superradiant dynamics}

\begin{figure}[htbp]
    \centering
    \includegraphics[width=0.5\textwidth]{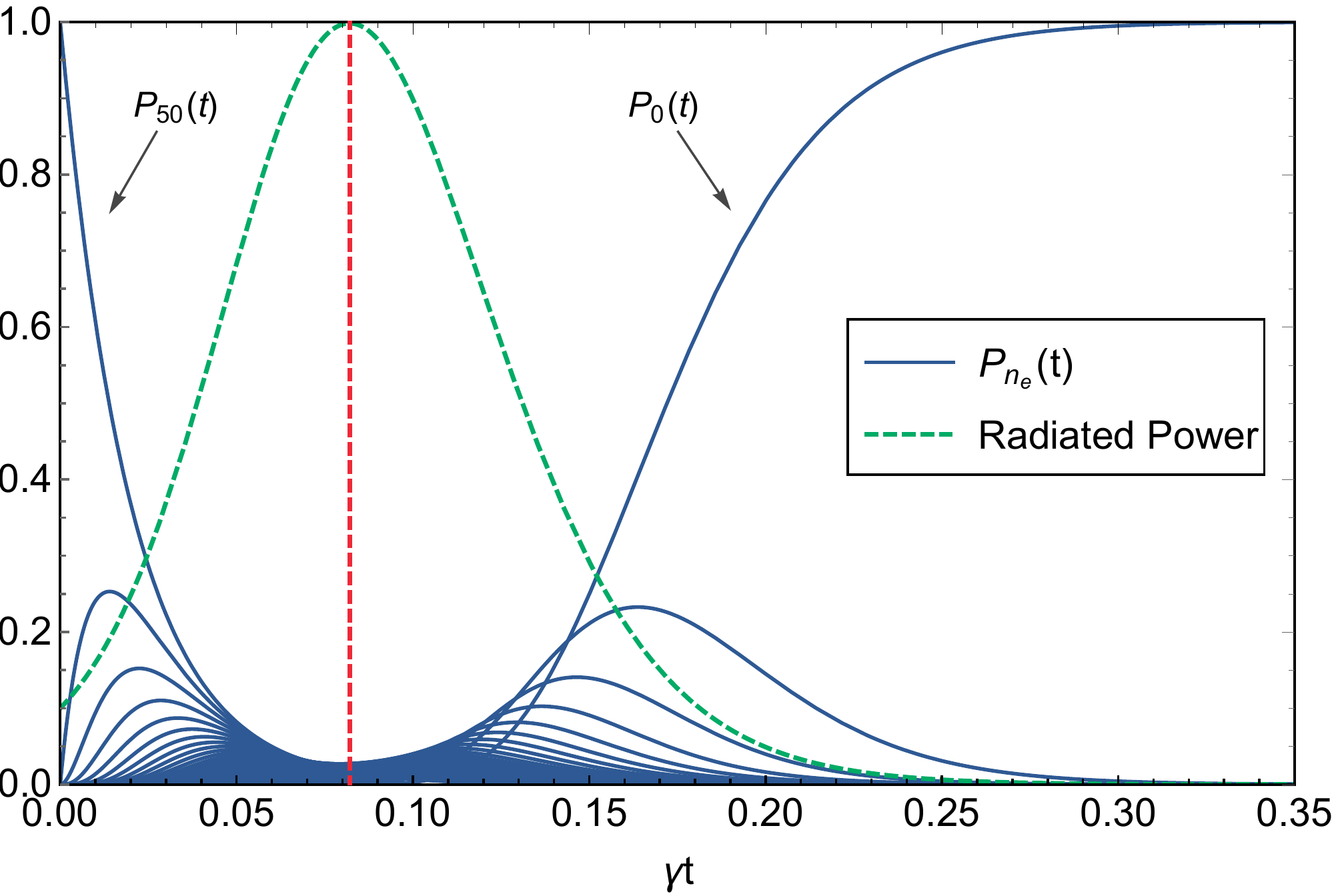}
    \caption{Population $P_{n_e}$ of each Dicke state (blue solid lines) as a function of $\gamma t$ during a superradiant emission for $N=50$ atoms. The green dashed line displays the  radiated power [Eq.~(\ref{eq:radiatedpower})] in normalized units (to fit the plot). The vertical red dashed line represents the time of maximum radiated power, $t^{\text{p}}_{\text{max}}$.}
    \label{fig:pop}
\end{figure}

Having characterized the GMCs of pure Dicke states, we now move on to analyze how the GMCs evolve during superradiant dynamics. 
We consider a system of $N$ identical two-level atoms with transition frequency $\omega$ between the ground $\ket{0}$ and excited $\ket{1}$ states. The atoms are assumed to interact indirectly through the electromagnetic vacuum, whose fluctuations cause them to decay and emit radiation. According to this model, if initially, all atoms in the sample are in the excited state, the state of the system for every time $t$ can be written as \cite{agarwal1974quantum} 
\begin{equation}
    \label{eq:superdensity2}
    \rho_N(t)=\sum_{n_e=0}^N P_{n_e}(t)\ket{N,n_e}\bra{N,n_e},
\end{equation}
with $P_{n_e}(t)$ being the population in each Dicke state over time. These populations evolve according to
\begin{equation}
    \frac{\partial P_{n_e}(t)}{\partial t}=\nu_{n_e+1} P_{n_e+1}(t)-\nu_{n_e} P_{n_e}(t),
\end{equation}
where $\nu_{n_e}=2\gamma n_e(N-n_e+1)$ and $\gamma$ is the atomic spontaneous decay rate. 
The evolution of $P_{n_e}(t)$ in a typical Dicke superrandiant process is shown in Fig.~\ref{fig:pop}. 
Here and henceforth, we always assume that the system is initially prepared with $P_{n_e = N}(0) = 1$.

The temporal signature of the superradiant emission is the radiated power $\mathcal{P}$, which is given by \cite{agarwal1974quantum}
\begin{equation}\label{eq:radiatedpower}
    \mathcal{P}(t) = 2\gamma\omega\sum\limits_{n_e=0}^N n_e\left(1+N-n_e\right)P_{n_e}(t).
\end{equation}
The radiated power is shown in  Fig. \ref{fig:pop} in a  green dashed line. 
It achieves its maximum value around the time $t^{\text{p}}_{\text{max}}=(N\gamma)^{-1}$. This reflects the fact that the cooperative effect of the atoms in superradiance allows the releasing of great amounts of radiation energy in a short period, making the intensity of the radiated power proportional to $N^2$ rather than $N$, as one would expect if the atoms were radiating incoherently. 

\subsection{GMCs in Dicke Superradiance}

\begin{figure}[htbp]
    \centering
    \includegraphics[width=0.5\textwidth]{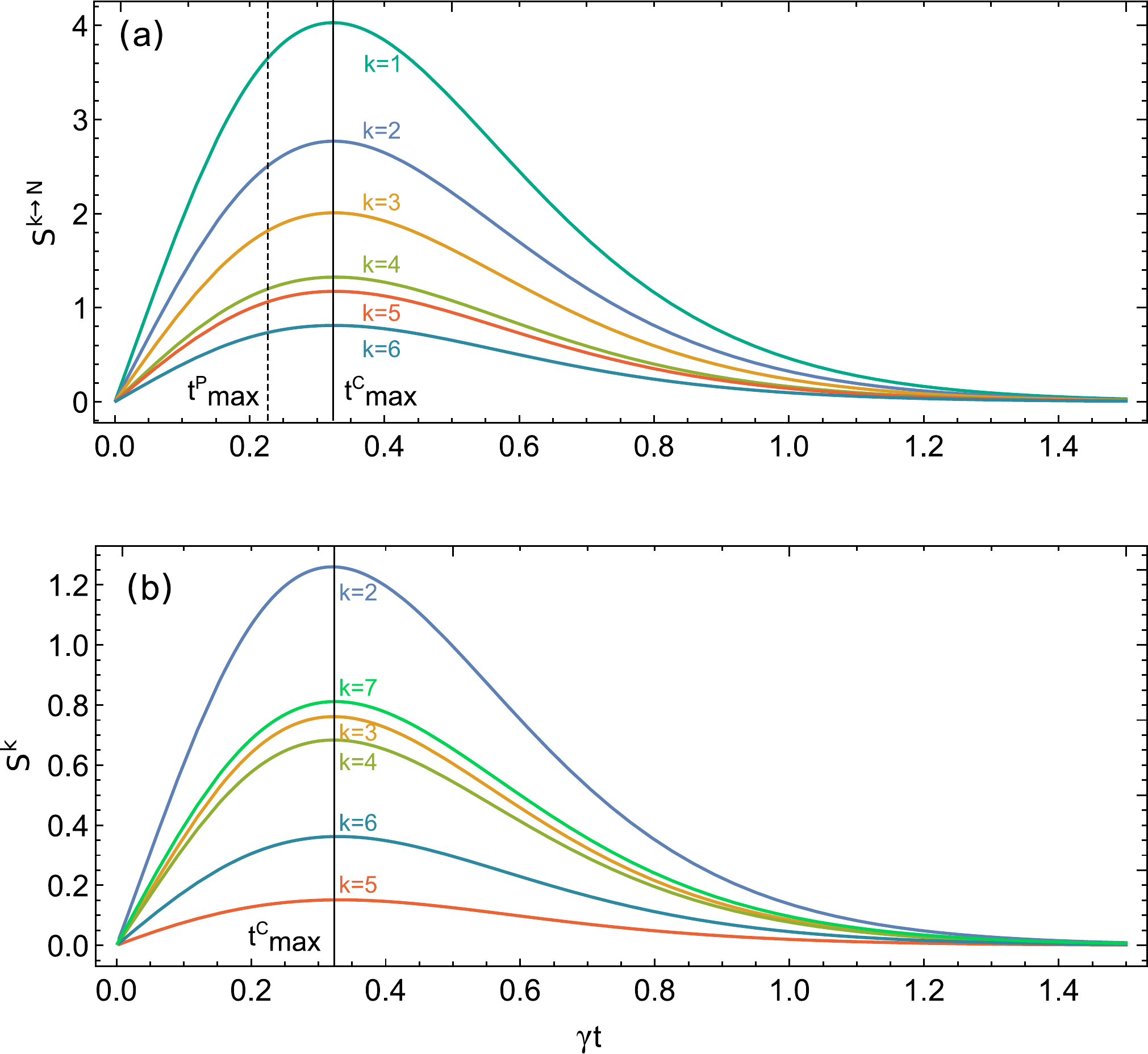}
    \caption{The GMCs (a) $S^{k\to N}$ and (b) $S^k$ during the superradiant dynamics with $N=7$. The vertical black dashed and solid lines represent the time in which the radiated power $t^{\text{P}}_{\text{max}}$ and the GMCs $t^{\text{C}}_{\text{max}}$ achieve their maximum values.}
    \label{fig:super7}
\end{figure}

We now move on to study the time evolution for the GMCs during the superradiant dynamics. We begin with small system sizes and present in Fig. \ref{fig:super7}  the GMCs for the superradiant system in the case $N=7$. The time of maximum correlation $t^{\text{c}}_{\text{max}}$ is universal for any value of $k$, depending only on the size of the system $N$. We observed numerically that the time of maximum correlation coincides with the time of maximum entropy of the system.
Since the initial $\ket{N,N}$ and final $\ket{N,0}$ states of the superradiance are product states, all correlations at these times are null, as expected. However, the GMCs become prominent around the time in which the radiated power is stronger.

Figure (\ref{fig:super7}a) shows that the GMCs of order higher than $k$ decrease as the size of the greatest partition increases, a behavior already presented by Dicke states, see Fig. (\ref{fig:dicke}a). The GMCs $S^k$  in Fig.~(\ref{fig:super7}b) are seen to contribute in a way that is out of order in $k$. In the case of $N=7$, the partitions with stronger genuine correlations are in decreasing order 2, 7, 3, 4, 6, and 5. 
Although the GMCs of order $k = 2$ dominate over other partitions, the genuine correlation between all atoms ($k = 7$) plays a significant role, even more than tripartite correlations, for instance. Similarly, GMCs between $k=6$ atoms is stronger than between $k=5$ atoms. These rather unintuitive results are a consequence of strong finite-size effects. 

\begin{figure}[htbp]
    \centering
    \includegraphics[width=0.5\textwidth]{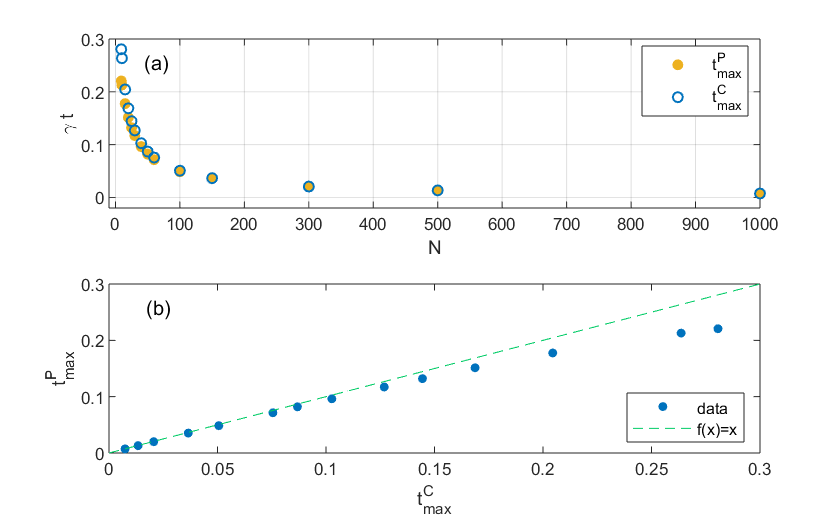}
    \caption{(a) Dimensionless time $\gamma t$ where the genuine multipartite correlations reach their maximum values $t^C_{\max}$ behaves similarly to the time in which the radiated power achieves its maximum value $t^P_{\max}$, i.e., both times are inversely proportional to the number of atoms $N$ in the sample. (b) The parametric plot of $t^P_{\max}(N)$ \textit{versus} $t^C_{\max}(N)$ shows that as $N$ increases these two times become equivalents.}
    \label{fig:times}
\end{figure}

From Fig. (\ref{fig:super7}a) it is possible to observe that the time in which the GMCs reach their maximum values, $t^C_{\max}$, occurs after the time of maximum radiated power  $t^P_{\max}$. 
The mismatch between them is a finite size effect and vanishes when $N\to \infty$, as shown in Fig. \ref{fig:times}. We find that $t^C_{\max}$ points out the temporal behavior of the superradiance phenomenon. Moreover, analogously to $t^P_{\max}$, one sees in Fig. (\ref{fig:times}a) that $t^C_{\max} \propto N^{-1}$. To explore more deeply the similarity between these two times, in Fig. (\ref{fig:times}b) we present the parametric plot $t^C_{\max}(N) \times t^P_{\max}(N)$, which shows that for larger values of $N$, i.e., smaller times, the time of maximum radiated power and the time of maximum correlations are equivalents.
\begin{figure}[htbp]
    \centering
    \includegraphics[width=0.5\textwidth]{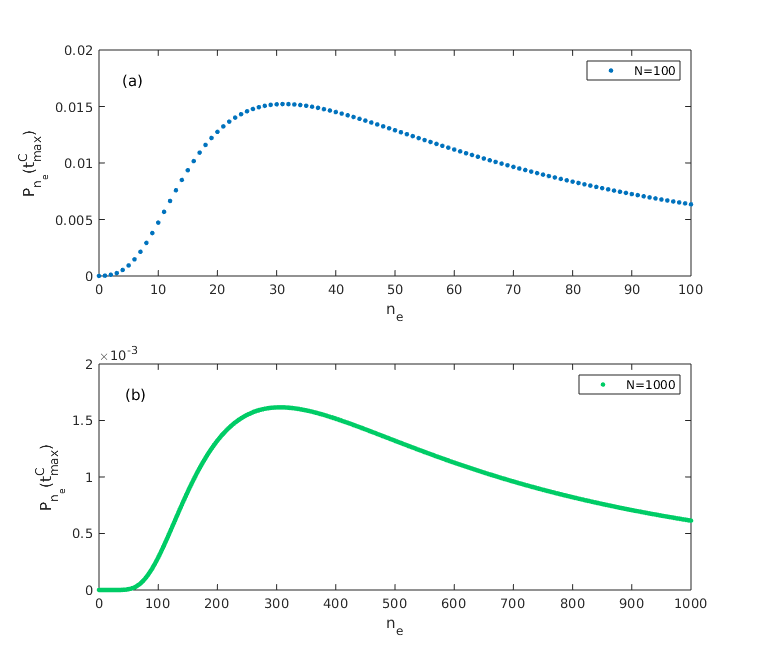}
    \caption{Population of Dicke states with (a) $N=100$ and (b) $N=1000$ in the superradiant state at the time of maximum correlation, $t_{\max}^C$.}
    \label{fig:pop_dicke}
\end{figure}

At the time of maximum correlations, the most populated Dicke states are those around $N/3$, as shown in Fig. (\ref{fig:pop_dicke}a) for $N=100$ and (\ref{fig:pop_dicke}b) for $N=1000$. For $N=1000$, the time of maximum emission of radiation $t^P_{max}$ and the time of maximum correlation $t^C_{max}$ are quite similar. This leads to the questions of what are the Dicke states which contribute significantly to the genuine multipartite correlations at $t^C_{max}$.

\begin{figure}[htbp]
    \centering
    \includegraphics[width=0.5\textwidth]{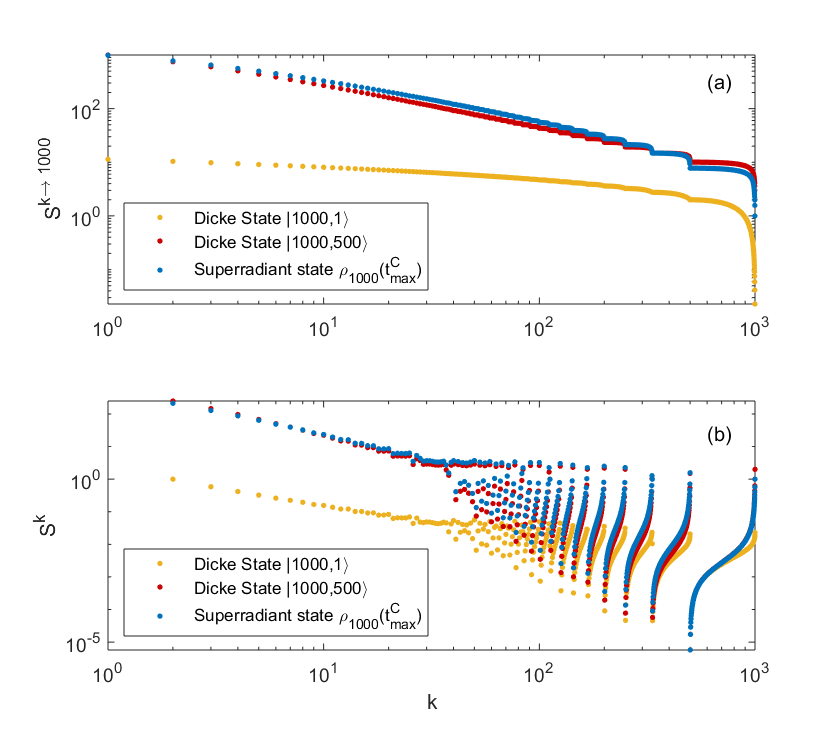}
    \caption{
    (a) $S^{k\to N}$ and (b) $S^k$ for the state at the time of maximum correlation, $t_\text{max}^C$ with $N = 1000$. For comparison we also present results for the pure $|N,1\rangle$ and $|N,N/2\rangle$ Dicke states.
    One can see that the GMCs stem predominantly from  $\ket{N,N/2}$.}
    \label{fig:super_cd}
\end{figure}

For this purpose, we plot in Fig. (\ref{fig:super_cd}) the GMCs for $1000$ atoms in two Dicke states, $\ket{1000,500}$ and $\ket{1000,1}$, and in the superradiant state at the moment in which the correlations are maximum, $\rho_{1000}\left( t^C_{\max} \right)$. We have chosen these two Dicke states provided that they bound the amount of correlations present in all Dicke states, as already shown in Fig. \ref{fig:dicke}. The results of $S^{k \to 1000}$ and $S^{k}$ in Figs. (\ref{fig:super_cd}a) and (\ref{fig:super_cd}b), respectively, show that the GMCs at $t^C_{\text{max}}$ are very similar to the most correlated Dicke state $\ket{1000, 500}$. However, increasing the size of the greatest partition $k$, the GMCs of the superradiant state becomes smaller than the GMCs of the Dicke state $\ket{N, N/2}$. We highlight that this result was obtained for $1000$ atoms in the sample.

\section{\label{sec:conc}Conclusions}

We computed the genuine $k$-partite correlations for Dicke states with an arbitrary number of excitations and extended it to the superradiant state. 
As a result, we found that genuine $k$-partite correlations for the superradiant state are not monotonic in the partition size $k$ and present strong finite-size effects. 
We showed that the time in which the genuine multipartite correlations reach a  maximum, $t_{\max}^C$, is universal for any partitioning $k$ and depends only on the size of the system. Numerically we observed that the time of maximum correlation is the same as the time of maximum entropy of the system.
We also observed that $t_{\max}^C$ is right after the time of maximum emission of radiation, $t_{\max}^P$. However, as the size of the system increases, both times tend to be the same.

\begin{acknowledgments}
We thank Renné Medeiros de Araújo and Thiago O. Maciel for fruitful discussions.
The authors acknowledge the financial support from the Brazilian funding agencies Coordenação de Aperfeiçoamento de Pessoal de Nível Superior (CAPES), Conselho Nacional de Desenvolvimento Científico e Tecnológico (CNPq), Fundação de Amparo à Pesquisa e Inovação do Estado de Santa Catarina (FAPESC) and the Instituto Nacional de Ciência e Tecnologia de Informação Quântica (INCT-IQ). 
GTL acknowledges the financial support from the São Paulo Research Foundation (FAPESP) under grant 2018/12813-0. 
\end{acknowledgments}

\appendix
\begin{widetext}
\section{Calculation of the GMCs for Dicke States}
\label{ap: Dicke}

The reduced density matrix of an arbitrary Dicke state $\ket{N,n_e}$ in Eq.~(\ref{eq:DickeState2}) is readily found to be 
\[
    \Tr_{N-k}(\ket{N,n_e}\bra{N,n_e}) = \sum_{i=0}^{n_e} \frac{\binom{k}{i}\binom{N-k}{n_e-i}}{\binom{N}{n_e}} \ket{k,i}\bra{k,i},
\]
where $\ket{k,i}$ is a Dicke state of $k$ particles and $i$ excitations. The binomial $\binom{n}{m}$ is defined to be zero whenever $m$ does not belong in the interval $n\geq m\geq 0$, which can happen in the equation above, for example, when $k$ is smaller than $n_e$.
As the Dicke states form an orthonormal basis for the subspace of completely symmetric states, the GMCs of order higher than $k$, Eq.~(\ref{eq:Skhigher_symm}), can be written as
\begin{align}
    S^{k\to N}(\ket{N,n_e})=&-\left\lfloor \frac{N}{k} \right\rfloor \sum_{i=0}^k h\left(\frac{\binom{k}{i}\binom{N-k}{n_e-i}}{\binom{N}{n_e}}\right)\nonumber\\
    &+(\delta_{N\text{mod}k,0}-1)\sum_{i=0}^{N\text{mod}k} h\left(\frac{\binom{N-\lfloor N/k \rfloor k}{i}\binom{\lfloor N/k \rfloor k}{n_e-i}}{\binom{N}{n_e}}\right),
    \label{eq:Dickekhigher}
\end{align}
where $\displaystyle h(x)=\lim_{y\to x}y \log(x)$.

The same reasoning applies to statistical mixtures of Dicke states, as in Eq.~(\ref{eq:superdensity2}). 
The reduced density matrix for the Dicke superradiant state is
\begin{equation}
    \rho_k=\sum_{j_e=0}^{k}\sum_{l_e=0}^{N-k} \frac{\binom{k}{j_e}\binom{N-k}{l_e}}{\binom{N}{j_e+l_e}}P_{j_e+l_e}\ket{k,j_e}\bra{k,j_e}.
    \label{eq:superreduced}
\end{equation}
Therefore, the GMCs of order higher than $k$ for superradiance are
\begin{align}
    S^{k\to N}(\rho_N)=&-\left\lfloor \frac{N}{k} \right\rfloor \sum_{j_e=0}^k h\left(\sum_{l_e=0}^{N-k} \frac{\binom{k}{j_e}\binom{N-k}{l_e}}{\binom{N}{j_e+l_e}}P_{j_e+l_e}\right)+\sum_{l_e=0}^N h[P_{l_e}(t)] \nonumber\\
    &+(\delta_{N\text{mod}k,0}-1)\sum_{j_e=0}^{N\text{mod}k} h\left(\sum_{l_e=0}^{\lfloor N/k \rfloor k} \frac{\binom{N-\lfloor N/k \rfloor k}{j_e}\binom{\lfloor N/k \rfloor k}{l_e}}{\binom{N}{j_e+l_e}}P_{j_e+l_e}\right).
    \label{eq:superhigher}
\end{align}
Albeit cumbersome, these formulas can readily be computed numerically. 

Large numbers can be a problem when computing the binomials in the equation above, as a numerical solution we define the binomial as

\begin{equation}
    \binom{n}{k}=\exp[\ln(\Gamma(n+1))-\ln(\Gamma(k+1))-\ln(\Gamma(n-k+1))].
\end{equation}
which allow us to have more stability in the computation.

\end{widetext}

\bibliographystyle{unsrt}
\bibliography{references.bib}

\begin{thebibliography}{10}

\bibitem{PhysRevA.62.062314}
W.~D\"ur, G.~Vidal, and J.~I. Cirac.
\newblock Three qubits can be entangled in two inequivalent ways.
\newblock {\em Phys. Rev. A}, 62:062314, Nov 2000.

\bibitem{coffman2000distributed}
Valerie Coffman, Joydip Kundu, and William~K Wootters.
\newblock Distributed entanglement.
\newblock {\em Physical Review A}, 61(5):052306, 2000.

\bibitem{greenberger1990bell}
Daniel~M Greenberger, Michael~A Horne, Abner Shimony, and Anton Zeilinger.
\newblock Bell’s theorem without inequalities.
\newblock {\em American Journal of Physics}, 58(12):1131--1143, 1990.

\bibitem{toth2005detecting}
G{\'e}za T{\'o}th and Otfried G{\"u}hne.
\newblock Detecting genuine multipartite entanglement with two local
  measurements.
\newblock {\em Physical review letters}, 94(6):060501, 2005.

\bibitem{walczak2010information}
Zbigniew Walczak.
\newblock Information-theoretic approach to the problem of detection of genuine
  multipartite classical correlations.
\newblock {\em Physics Letters A}, 374(39):3999--4002, 2010.

\bibitem{grudka2008note}
Andrzej Grudka, Michal Horodecki, Pawel Horodecki, and Ryszard Horodecki.
\newblock Note on genuine multipartite classical correlations.
\newblock {\em arXiv preprint arXiv:0802.1633}, 2008.

\bibitem{giorgi2013genuine}
Gian~Luca Giorgi and Thomas Busch.
\newblock Genuine correlations in finite-size spin systems.
\newblock {\em International Journal of Modern Physics B}, 27(01n03):1345034,
  2013.

\bibitem{zhang2012two}
Zhanjun Zhang, Biaoliang Ye, and Shao-Ming Fei.
\newblock Two different definitions on genuine quantum and classical
  correlations in multipartite systems do not coincide in general.
\newblock {\em arXiv preprint arXiv:1206.0221}, 2012.

\bibitem{bai2013exploring}
Yan-Kui Bai, Na~Zhang, Ming-Yong Ye, and ZD~Wang.
\newblock Exploring multipartite quantum correlations with the square of
  quantum discord.
\newblock {\em Physical Review A}, 88(1):012123, 2013.

\bibitem{grimsmo2012dynamics}
Arne~L Grimsmo, Scott Parkins, and Bo-Sture~K Skagerstam.
\newblock Dynamics of genuine multipartite correlations in open quantum
  systems.
\newblock {\em Physical Review A}, 86(2):022310, 2012.

\bibitem{li2012detecting}
Bo~Li, Leong~Chuan Kwek, and Heng Fan.
\newblock Detecting genuine multipartite correlations in terms of the rank of
  coefficient matrix.
\newblock {\em Journal of Physics A: Mathematical and Theoretical},
  45(50):505301, 2012.

\bibitem{giorgi2011genuine}
Gian~Luca Giorgi, Bruno Bellomo, Fernando Galve, and Roberta Zambrini.
\newblock Genuine quantum and classical correlations in multipartite systems.
\newblock {\em Physical review letters}, 107(19):190501, 2011.

\bibitem{giorgi2013erratum}
Gian~Luca Giorgi, Bruno Bellomo, Fernando Galve, and Roberta Zambrini.
\newblock Erratum: Genuine quantum and classical correlations in multipartite
  systems [phys. rev. lett. 107, 190501 (2011)].
\newblock {\em Physical Review Letters}, 110(13):139904, 2013.

\bibitem{maziero2012genuine}
Jonas Maziero and F{\'a}bio~M Zimmer.
\newblock Genuine multipartite system-environment correlations in decoherent
  dynamics.
\newblock {\em Physical Review A}, 86(4):042121, 2012.

\bibitem{aolita2012fully}
Leandro Aolita, Rodrigo Gallego, Ad{\'a}n Cabello, and Antonio Ac{\'\i}n.
\newblock Fully nonlocal, monogamous, and random genuinely multipartite quantum
  correlations.
\newblock {\em Physical review letters}, 108(10):100401, 2012.

\bibitem{novo2013genuine}
Leonardo Novo, Tobias Moroder, and Otfried G{\"u}hne.
\newblock Genuine multiparticle entanglement of permutationally invariant
  states.
\newblock {\em Physical Review A}, 88(1):012305, 2013.

\bibitem{mendoncca2015heuristic}
Paulo~EMF Mendon{\c{c}}a, Marcelo~A Marchiolli, and Gerard~J Milburn.
\newblock Heuristic for estimation of multiqubit genuine multipartite
  entanglement.
\newblock {\em International Journal of Quantum Information}, 13(03):1550023,
  2015.

\bibitem{girolami2017quantifying}
Davide Girolami, Tommaso Tufarelli, and Cristian~E. Susa.
\newblock Quantifying genuine multipartite correlations and their pattern
  complexity.
\newblock {\em Phys. Rev. Lett.}, 119:140505, Oct 2017.

\bibitem{modi2010unified}
Kavan Modi, Tomasz Paterek, Wonmin Son, Vlatko Vedral, and Mark Williamson.
\newblock Unified view of quantum and classical correlations.
\newblock {\em Physical review letters}, 104(8):080501, 2010.

\bibitem{bennett2011postulates}
Charles~H Bennett, Andrzej Grudka, Micha{\l} Horodecki, Pawe{\l} Horodecki, and
  Ryszard Horodecki.
\newblock Postulates for measures of genuine multipartite correlations.
\newblock {\em Physical Review A}, 83(1):012312, 2011.

\bibitem{susa2018weaving}
Cristian~E Susa and Davide Girolami.
\newblock Weaving and neural complexity in symmetric quantum states.
\newblock {\em Optics Communications}, 413:157--161, 2018.

\bibitem{i2016characterizing}
Jordi~Tura i~Brugu{\'e}s.
\newblock {\em Characterizing entanglement and quantum correlations constrained
  by symmetry}.
\newblock Springer, 2016.

\bibitem{dicke1954coherence}
Robert~H Dicke.
\newblock Coherence in spontaneous radiation processes.
\newblock {\em Physical Review}, 93(1):99, 1954.

\bibitem{gross1982superradiance}
Michel Gross and Serge Haroche.
\newblock Superradiance: An essay on the theory of collective spontaneous
  emission.
\newblock {\em Physics reports}, 93(5):301--396, 1982.

\bibitem{Bergmann_2013}
Marcel Bergmann and Otfried Gühne.
\newblock Entanglement criteria for {Dicke} states.
\newblock {\em Journal of Physics A: Mathematical and Theoretical},
  46(38):385304, 2013.

\bibitem{moreno2018all}
MGM Moreno and Fernando Parisio.
\newblock All bipartitions of arbitrary {Dicke} states.
\newblock {\em arXiv preprint arXiv:1801.00762}, 2018.

\bibitem{wolfe2014certifying}
Elie Wolfe and SF~Yelin.
\newblock Certifying separability in symmetric mixed states of n qubits, and
  superradiance.
\newblock {\em Physical review letters}, 112(14):140402, 2014.

\bibitem{yu2016separability}
Nengkun Yu.
\newblock Separability of a mixture of {Dicke} states.
\newblock {\em Physical Review A}, 94(6):060101, 2016.

\bibitem{tura2018separability}
Jordi Tura, Albert Aloy, Ruben Quesada, Maciej Lewenstein, and Anna Sanpera.
\newblock Separability of diagonal symmetric states: a quadratic conic
  optimization problem.
\newblock {\em Quantum}, 2:45, 2018.

\bibitem{santos2016}
Eduardo~M dos Santos and Eduardo~I Duzzioni.
\newblock Elucidating {Dicke} superradiance by quantum uncertainty.
\newblock {\em Physical Review A}, 94(2):023819, 2016.

\bibitem{szalay2015}
Szil\'ard Szalay.
\newblock Multipartite entanglement measures.
\newblock {\em Phys. Rev. A}, 92:042329, Oct 2015.

\bibitem{agarwal1974quantum}
Girish~S Agarwal.
\newblock Quantum statistical theories of spontaneous emission and their
  relation to other approaches.
\newblock In {\em Quantum Optics}, pages 1--128. Springer, 1974.

\end{thebibliography}

\end{document}